# A gate-variable spin current demultiplexer based on graphene

Li Su[1,2,3,†], Xiaoyang Lin[1,†], Youguang Zhang[1], Arnaud Bournel[2,3], Yue Zhang[1], Jacques-Olivier Klein[2,3], Weisheng Zhao[1,*], and Albert Fert[1,4]

[1] *Fert Beijing Research Institute, School of Electrical and Information Engineering, BDBC, Beihang University, Beijing, 100191, China*
[2] *Institut d'Electronique Fondamentale, Univ. Paris-Sud, Univ.Paris-Saclay, F-91405 Orsay, France*
[3] *UMR 8622, CNRS, F-91405 Orsay, France*
[4] *Unité Mixte de Physique CNRS-Thales, F-91767 Palaiseau, France*

† These two authors contribute equally.

* E-mail: weisheng.zhao@buaa.edu.cn

**Spintronics, which utilizes spin as information carrier, is a promising solution for nonvolatile memory and low-power computing in the post-Moore era. An important challenge is to realize long distance spin transport, together with efficient manipulation of spin current for novel logic-processing applications. Here, we describe a gate-variable spin current demultiplexer (GSDM) based on graphene, serving as a fundamental building block of reconfigurable spin current logic circuits. The concept relies on electrical gating of carrier density dependent conductivity and spin diffusion length in graphene. As a demo, GSDM is realized for both single-layer and bilayer graphene. The distribution and propagation of spin current in the two branches of GSDM depend on spin**

**relaxation characteristics of graphene. Compared with Elliot-Yafet spin relaxation mechanism, D'yakonov-Perel mechanism results in more appreciable gate-tuning performance. These unique features of GSDM would give rise to abundant spin logic applications, such as on-chip spin current modulators and reconfigurable spin logic circuits.**

Spintronics employs spin currents instead of charge currents for information storage, transfer, and processing, which has led to paradigmatic advances with, for example, nonvolatile memories and low-power logic circuits[1–3]. Power-efficient spin current generation, manipulation and detection have been extensively studied to implement practical spin logic applications, adopting metal or conventional semiconductor materials as the spin transport channel[4,5]. In recent years, graphene has attracted considerable attentions thanks to its superior features including high electronic mobility, weak spin-orbit coupling and hyperfine interactions[6,7,8]. As a promising spin transport channel, graphene achieves the longest spin diffusion length (SDL) at room temperature[9,10,11,12,13]. Further efforts have been made to promote graphene-based spin logic devices, such as spin transistors[14], magneto-logic devices[15] and all-spin logic devices[16,17]. However, most of them suffer from large dynamic power consumption during magnetic field operation or frequent data transmission between electrical and magnetic states for pipeline computing.

In this study, we describe a gate-variable spin current demultiplexer (GSDM) based on graphene, which could serve as a fundamental building block of reconfigurable spin current-based logic. We model the simplest GSDM with a Y-shaped structure in which the injection and propagation of spin currents in the two branches are controlled by reacting with gate voltages on the relative conductivities and spin relaxation times of graphene. The result of the study depends on the type of spin relaxation. Two mechanisms of spin relaxation are generally considered in graphene, Elliot-Yafet (EY)[18] and Dyakonov-Perel (DP)[19,20]. The EY mechanism is generally predominant in single layer graphene (SLG) with, however, some experimental results accounted for by a mixing EY and DP terms[11,21–26]. In bilayer graphene (BLG), the DP mechanism is generally found to be predominant[24,25,27,28]. We discuss the operation of GSDM in the two typical cases, i.e., SLG with EY mechanism and BLG with DP mechanism. In particular, BLG with DP mechanism is expected to achieve a more efficient voltage-control of spin current in GSDM. In addition, such a difference in spin current transport implies the possibility to figure out spin relaxation mechanisms in graphene by using a structure similar to our GSDM. Furthermore, the unique features of GSDM make it possible to design on-chip spin current modulators and reconfigurable spin logic circuits.

## Results

**Concept of graphene-based GSDM.**

The simplest GSDM, a Y-shaped graphene channel with gate voltages applied on both branches ($V_{G,1}$ and $V_{G,2}$) is shown in Fig. 1a; the top/back gate or both the gates can be designed for practical applications. Pure spin currents are injected into the two branches through diffusion due to spin accumulation $\Delta\mu_s$ ($x = 0$) at the bifurcation.

This spin accumulation can be created by spin injection upstream as in a classical nonlocal spin valve[7,9,24,29]. The spin current density in branch $i$ at distance $x$ from the bifurcation and for infinite length branches can be expressed as[30,31]

$$j_{s,i}(x,V_{G,i}) = \frac{\sigma_i}{e\lambda_{s,i}} \Delta\mu_s(x=0)e^{-x/\lambda_{s,i}}, \tag{1}$$

where $\sigma_i$ is the conductivity and $\lambda_{s,i}$ is the SDL of branch $i$. $\sigma_i$ and $\lambda_{s,i}$ can be controlled using the gate voltage $V_{G,i}$ via the corresponding carrier density $n_i$. The ratio $\sigma_i/\lambda_{s,i}$ dominates the distribution of spin current $j(x = 0, V_{G,i})$ at the bifurcation ($x = 0$), and the exponential decay $e^{-x/\lambda_{s,i}}$ describes the propagation of the spin current at distance $x$ in the branches. The gate voltage control of the carrier density $n_i$, conductivity $\sigma_i(n_i)$, and SDL $\lambda_{s,i}(n_i)$ can be used to direct the spin current to the left or right branch, or equally to both branches. Notably, the spin current flows can be reconfigured in a circuit without using any magnetic field. This is favorable to build up low-power spintronic circuits and energy-efficient architectures.

**Dependence of conductivity and spin diffusion length on carrier density in SLG and BLG.**

The conductivity σ can be expressed as $\sigma = \sigma(n) + \sigma_{min}$, where the first term is proportional to the global carrier density $n$ and mobility $\mu$, while $\sigma_{min}$ is the $n$-independent minimum conductivity related to the residual carrier density induced by inhomogeneous charge distribution in the small $n$ limit. We consider only the regime with sufficiently large $n$, typically $n > 10^{12}\,cm^{-2}$, in which $\sigma_{min}$ can be neglected and σ can be expressed as

$$\sigma = ne\mu \qquad (2)$$

In Eq. (2), for relatively large $n$, the mobility $\mu$ turns out to be approximately independent of $n$ in most experiments, for example, in Fig. 2 in the study by Tan et al[33]. At room temperature, we used a typical high mobility of 15,000 cm$^2$/Vs for SLG[32] and a smaller mobility 10,000 cm$^2$/Vs for BLG[27]. The second parameter in Eq. (1) is the SDL

$$\lambda_s = v_F (\tau_p \tau_s/2)^{1/2} \qquad (3)$$

where $v_F$ is the Fermi velocity, $\tau_p$ is the momentum relaxation time and $\tau_s$ is the spin relaxation time.

**SLG:** From the Einstein relation for electrons with linear dispersion in SLG, the conductivity is proportional to $\tau_{p,\text{SLG}}(n)^{1/2}$, where $\tau_{p,\text{SLG}}$ is the momentum-relaxation time. Consequently, the approximate $n$-independent mobility obtained in experiments corresponds to $\tau_{p,\text{SLG}} \sim (n)^{1/2}$ of Eq. (2) (see the complete expression of $\tau_{p,\text{SLG}}$ and other parameters in Table 1). The variation of the spin-relaxation time on $n$ depends on the type of spin-relaxation mechanism, that is, the EY[18] mechanism related to spin-orbit energy $\Delta_{\text{EY}}$ or the DP[19,20] mechanism related to spin-orbit energy $\Delta_{\text{DP}}$.

For the EY mechanism[33], $\tau_{s,\text{EY}} \approx \dfrac{E_F^2}{\Delta_{\text{EY}}^2}\tau_p$, with $E_F^2 \sim k_F^2 \sim n$ for Dirac electrons. This results in $\tau_{s,\text{EY}} \sim (n)^{3/2}$. Therefore, from Eq.(3) with $\tau_{p,\text{SLG}} \sim (n)^{1/2}$ we get

$$\lambda_{s,\text{SLG,EY}} \sim n \qquad (4)$$

and, based on Eqs. (2) and (4), the prefactor of Eq.(1), $\sigma/\lambda_s$, is independent of $n$. In SLG with EY, different gate voltages, resulting in different values of $n$ in the left and right channels cannot generate different injected spin currents at $x = 0$; however, the dependence of $\lambda_{s,\text{SLG,EY}}$ on $n$ causes these spin currents to decay differently in the two branches.

For DP relaxation, $\tau_s$ is inversely proportional to $\tau_p$, $\tau_{s,DP} \approx \dfrac{\hbar^2}{4\Delta_{DP}^2} \dfrac{1}{\tau_p}$ [34], and $\lambda_{s,\text{SLG,DP}}$ of Eq. (3) is independent of $n$, whereas, from Eq. (2)

$$\sigma/\lambda_{s,\text{SLG,DP}} \sim n \tag{5}$$

Therefore, for SLG with DP, the injected spin currents are different at $x = 0$ (in proportion to the different values of $n$ in the left and right channels) and decay in a similar manner as a function of $x$.

**BLG:** For BLG and quadratic dispersion, the mobility independent of $n$ is equivalent to a momentum-relaxation time independent of $n$. For EY spin relaxation with $\tau_{s,EY} \approx \dfrac{E_F^2}{\Delta_{EY}^2}\tau_p$ and $E_F \sim n$, $\tau_{s,\text{EY}}$ is proportional to $n^2$ and

$$\lambda_{s,\text{BLG,EY}} \sim n \tag{6}$$

From Eqs. (2) and (6), $\sigma/\lambda_s$ does not depend on $n$. The different values of $n$ in the two branches do not lead to different injected spin currents at $x = 0$; however, as $\lambda_{s,\text{BLG,EY}}$ is proportional to $n$, the spin currents decay differently in the two branches.

Finally, for BLG with the DP mechanism, $\tau_s \tau_p$ and $\lambda_s$ are independent of $n$ and

$$\sigma/\lambda_{s,\text{BLG,DP}} \sim n \tag{7}$$

Similar to SLG with DP, for BLG with DP, the injected spin currents are different at $x = 0$ (in proportion to $n$ at the left and right channels) and decay in a similar manner as a function of $x$.

**Controlling the distribution of spin currents through gate voltage**

The carrier density $n_i$ in branch $i$ of the device can be controlled using the gate voltage $V_{G,i}$ according to the expression

$$n_i(V_{G,i}) = \varepsilon_0 \varepsilon_G (V_{G,i} - V_{G,0})/(t_G e), \qquad (8)$$

where $\varepsilon_0$ is the electric constant, $\varepsilon_G$ is the dielectric constant of the gates, $t_G$ is the dielectric thickness, and $V_{G,0}$ is the voltage required at the charge neutrality point. We calculated the ratio $j_{s,1}(x, V_{G,1})/j_{s,2}(x, V_{G,2})$ between the spin currents in the two branches of our device for different values of $v_{G,i} = (V_{G,i} - V_{G,0})$. The calculation is performed for two of the four cases previously discussed, SLG with EY spin relaxation (equally injected currents at bifurcation but different propagations) and BLG with DP spin relaxation (different injected currents at right and left channels with similar decays). Actually, in some experimental results, such as in those by Zomer et al.[11], the relaxation is considered by mixing EY and DP relaxation mechanisms, and is modeled in a straightforward extension of our approach.

Figs. 2 and 3 summarize the proportion results of the spin currents derived from Eq.(1) in the two branches by using the dependences of $\sigma_i$ and $\lambda_{s,i}$ on $n_i$ and $v_{G,i}$ expected from Eqs.(2)–(8). For SLG with EY relaxation, the spin currents injected into the two branches at bifurcation are equal for any applied voltage (Fig. 2a); however, their voltage-dependent propagation causes large differences when $x$ becomes of the order of or larger than the SDL in the branch with shorter SDLs (Fig. 2c). For example, by calculating the SDLs from the corresponding expression in Table 1 with $\Delta_{EY} = 10$ meV in the typical range derived from experiences in Refs. 11 and 22, we obtain 3.4 μm for $v_{G,1}$ = 5 V in branch 1 and 0.6 μm for $v_{G,2}$ = 0.5 V in branch 2, so that at $x$ = 1 μm smaller (larger) than the SDL in branch 1 (2), the spin current is strongly reduced in branch 2 but practically not reduced in branch 1 ($j_{s,1}/j_{s,2}$ = 4.5; Fig. 2c). At $x$ close to or larger the SDL in branch 1, $j_{s,1}/j_{s,2}$ would be even larger but this would be compensated by a small current reduction in branch 1.

For BLG with DP relaxation (Figs. 2b and 2d), the ratio between the spin currents can be as large as approximately 7 when the voltages are 0.5 V and 5 V on the opposite branches (Fig. 2b), and it does not depend on the distance from the bifurcation (Fig. 2d) because the damping is independent of the gate voltage. The SDL calculated from the corresponding equation in Table 1 using a typical experimental spin-orbit value for BLG, that is, $\Delta_{DP}$ = 0.14 meV (see Ref. 28), is 1.7 μm. This indicates that the different respective values of the spin currents at the start of the bifurcation remain unchanged without significant damping at distance $x$ in the micrometer range. From the corresponding expression in Table 1, the general behavior described earlier is illustrated in color scale for different couples of voltages in Fig. 3.

**Discussion and conclusion**

We have shown that tuning the conductivity and spin diffusion length by gate voltage in graphene can be used to control the distribution of spin currents in a spintronic logic circuit with graphene-based GSDM. We used two examples of Y-shaped GSDM to show how the spin current can be directed either predominantly into the left or right branch or equally toward both. Different behaviors are obtained with EY and DP spin relaxations. The BLG with DP relaxation, in which a strong contrast can be established between the currents in the two branches after the bifurcation and maintained over several micrometers in our simulations, is a more adapted logic circuit with typical distances in micrometers between logic gates. In addition, the marked difference between the operation of GSDM for DP and EY relaxation mechanisms indicates a feasible strategy to characterize the spin relaxation mechanisms.

**Methods**

We investigated spin current and accumulation based on the model, considering the spin-dependent and one-dimensional drift-diffusion theory[30, 31]. The electrical current density gives $j_{\uparrow,\downarrow} = -(\sigma_{\uparrow,\downarrow}/e)\nabla\mu_{\uparrow,\downarrow}$, and spin current is defined as $j_s = j_\uparrow - j_\downarrow$. Thus, the spin current density of spin up or down electrons is calculated as $j_s = -(\sigma/e)\nabla\mu_s$ in nonmagnetic spin channel material with $\sigma_\uparrow = \sigma_\downarrow$, where $\mu_s = \mu_\uparrow - \mu_\downarrow$ is the spin chemical potential. In addition, the decay of $\mu_s$, away from the interface, is characterized by the corresponding SDL $\mu_s(x) = \mu_s(0)e^{-x/\lambda_s}$. The spin current density of different channels is obtained consequently, as illustrated in Eq. (1).

The properties of charge transport are represented by the conductivity and momentum-scattering time of graphene. For SLG, the momentum scattering time $\tau_{p,SLG}$ is derived from the Einstein relation, $\sigma = e^2 \upsilon(E) D_c$. The energy is unknown here but the density of states related to the energy $E$ in graphene is given by $\upsilon(E) = \frac{g_v g_s |E|}{2\pi(\hbar v_F)^2}$ [8], which can further be used to calculate the carrier density $n(E) = \int_0^E \upsilon(E)dE = g_s g_v E^2 / [4\pi(\hbar v_F)^2]$. Based on the aforementioned equations, the charge diffusion coefficient is derived as $D_c = \dfrac{\sigma}{e^2} \dfrac{\sqrt{\pi}\hbar v_F}{\sqrt{g_v g_s n(E)}}$. Next, we substitute $D_c$ in the equation $D_c = v_F^2 \tau_{p,SLG} / 2$ to deduce the *n*-dependent momentum scattering time (Table 1). Moreover, the momentum scattering time of BLG $\tau_{p,BLG}$ can be derived similarly. Further, assuming that the spin diffusion coefficient $D_s \approx D_c$, SDL is calculated by $\lambda_s = \sqrt{D_s \tau_s}$ with the spin diffusion time $\tau_s$.

The Fermi level of graphene can simply be tuned by applying a voltage gate. This results in gate-variable carrier density in graphene. In the case of an electrically gated graphene, the gate $V_G$ would generate an electric field that is experienced by the graphene, that is, $E_G(V_G) = -\varepsilon_G (V_G - V_{G,0}) / t_G$. The variable $\varepsilon_G$ is the dielectric constant of the gate, $t_G$ is the dielectric thickness, and $V_{G,0}$ is the position of the charge neutrality point. Thus, the carrier density can be given as $n(V_G) = \varepsilon_0 E_G(V_G) / e$, where $\varepsilon_0$ is the electric constant.

## ACKNOWLEDGMENT

W.S.Z. thanks the financial support of the International Collaboration 111 Project B16001 from the Ministries of Education and Foreign Experts, Beijing Natural Science Foundation (4162039) and the National Natural Science Foundation of China (Grant No. 61504006 and No.61571023). This work was also supported by the Innovation Foundation of Beihang University (BUAA) for PhD Graduates.

## AUTHOR CONTRIBUTIONS

W.S.Z. and A.F. coordinated the proposal and supervised the project. L.S., X.Y.L., Y.Z., W.S.Z., and A.F. designed the device structure and its function. L.S. developed the model and performed the calculations. L.S., X.Y.L., W.S.Z., Y.G.Z, J.O.K., and A.F. analyzed the data. L.S., X.Y.L., W.S.Z., A.B., and A.F. edited the manuscript. All authors reviewed and commented the manuscript.

## Additional information

**Supplementary Information** accompanies this paper at

http://www.nature.com/naturenanotechnology.

## COMPETING FINANCIAL INTERESTS STATEMENT

The authors declare no competing financial interests.

**FIGURES AND FIGURE LEGENDS**

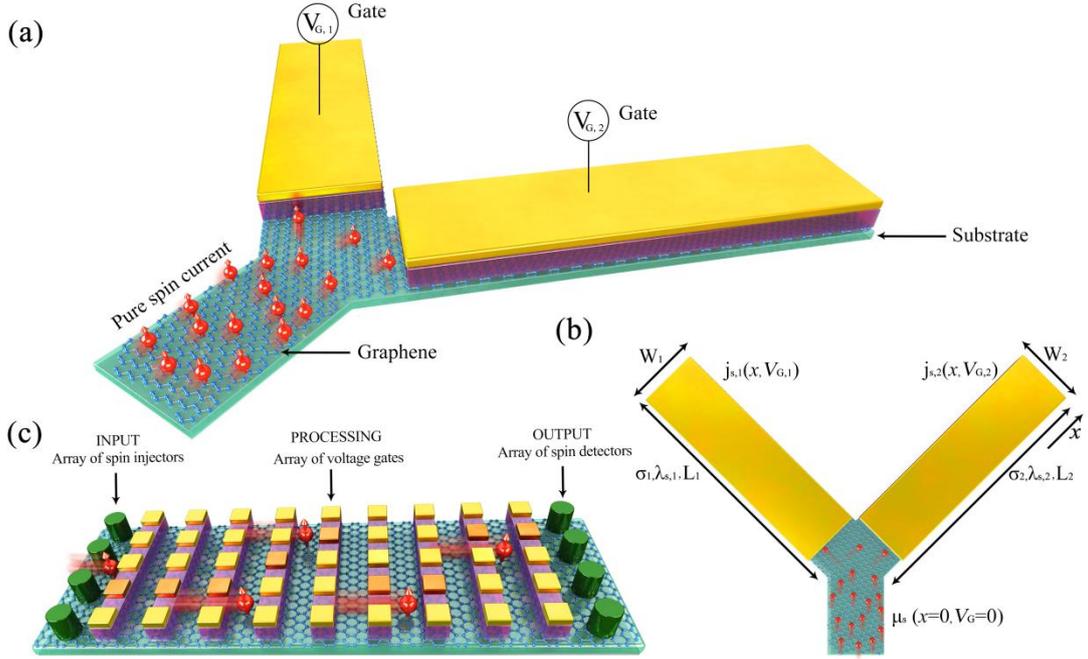

**Figure 1 | Schematics of graphene-based GSDM for reconfigurable spin logic circuit.** (**a**) A Y-shaped graphene-based GSDM with gate $V_{G,i}$ for voltage-control of spin currents in reconfigurable spin logic. An additional classical nonlocal spin valve structure to generate and detect pure spin currents is not shown in detail. (**b**) Top view of the Y-shaped GSDM with voltage-control; significant parameters, such as material characteristics ($\sigma_i$, $\lambda_{s,i}$); and geometrical parameters (width $W_i$ and distance $x$). The channel length is supposed to be much longer than the SDL, that is, $L_i \gg \lambda_{s,i}$, and the structure is symmetric, that is, $L_1 = L_2$ and $W_1 = W_2$. (**c**) Reconfigurable spin logic circuit based on gate control of spin current flow, with inputs, processing and outputs by spin injection, voltage and spin detection gates.

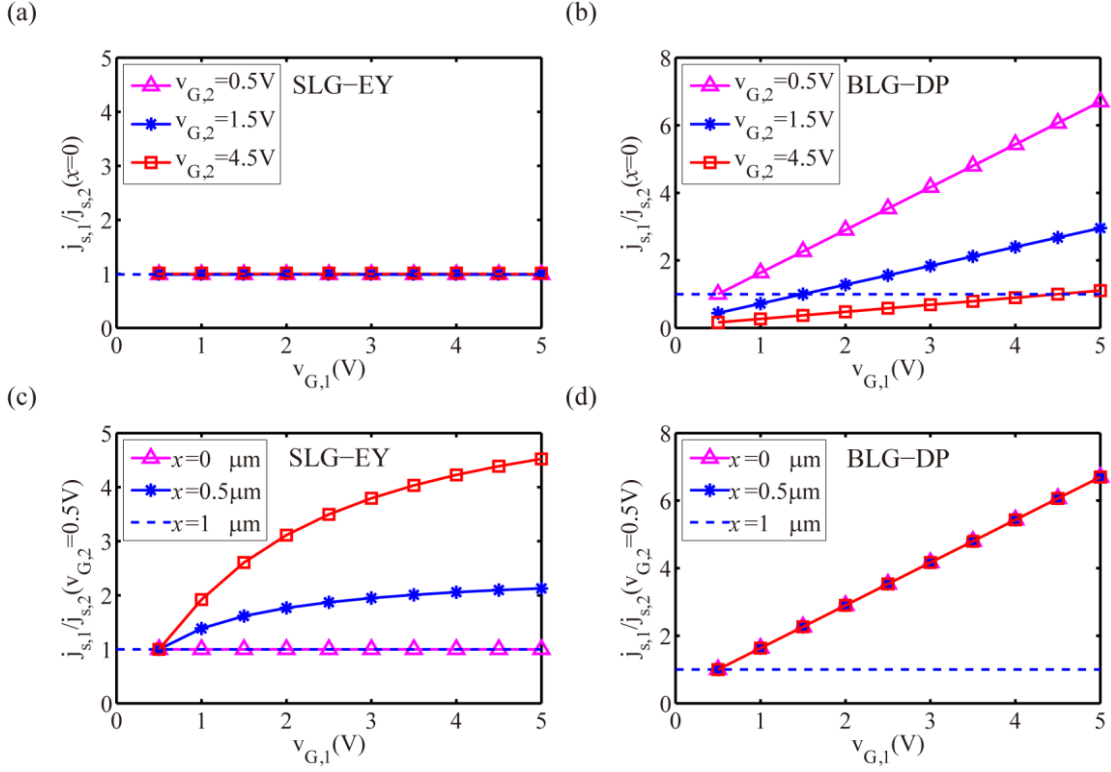

**Figure 2 | Spin current ratio of Y-shaped GSDM.** The ratio between the spin current in the two branches $j_{s,1}/j_{s,2}$ is plotted as a function of $v_{G,1}$ for different values of $v_{G,2}$ or a different distance $x$ from the bifurcation of SLG with EY mechanism (SLG-EY) on the left and BLG with DP mechanism (BLG-DP) on the right. The blue dashed line represents $j_{s,1}/j_{s,2}=1$. **(a)** and **(b)** $j_{s,1}/j_{s,2}$ at the bifurcation ($x = 0$) as the function of voltage $v_{G,1} = (V_{G,1} - V_{G,0})$ for $v_{G,2} = (V_{G,2} - V_{G,0}) = 0.5$, 1.5, or 4.5 V. **(c)** and **(d)** $j_{s,1}/j_{s,2}$ at distances $x = 0$, 0.5, and 1 μm as a function of voltage $v_{G,1}$ when $v_{G,2}$ is fixed at 0.5 V. The calculations are performed with $\varepsilon_G \approx 3.9$ and $t_G = 50$ nm for the dielectric gate.

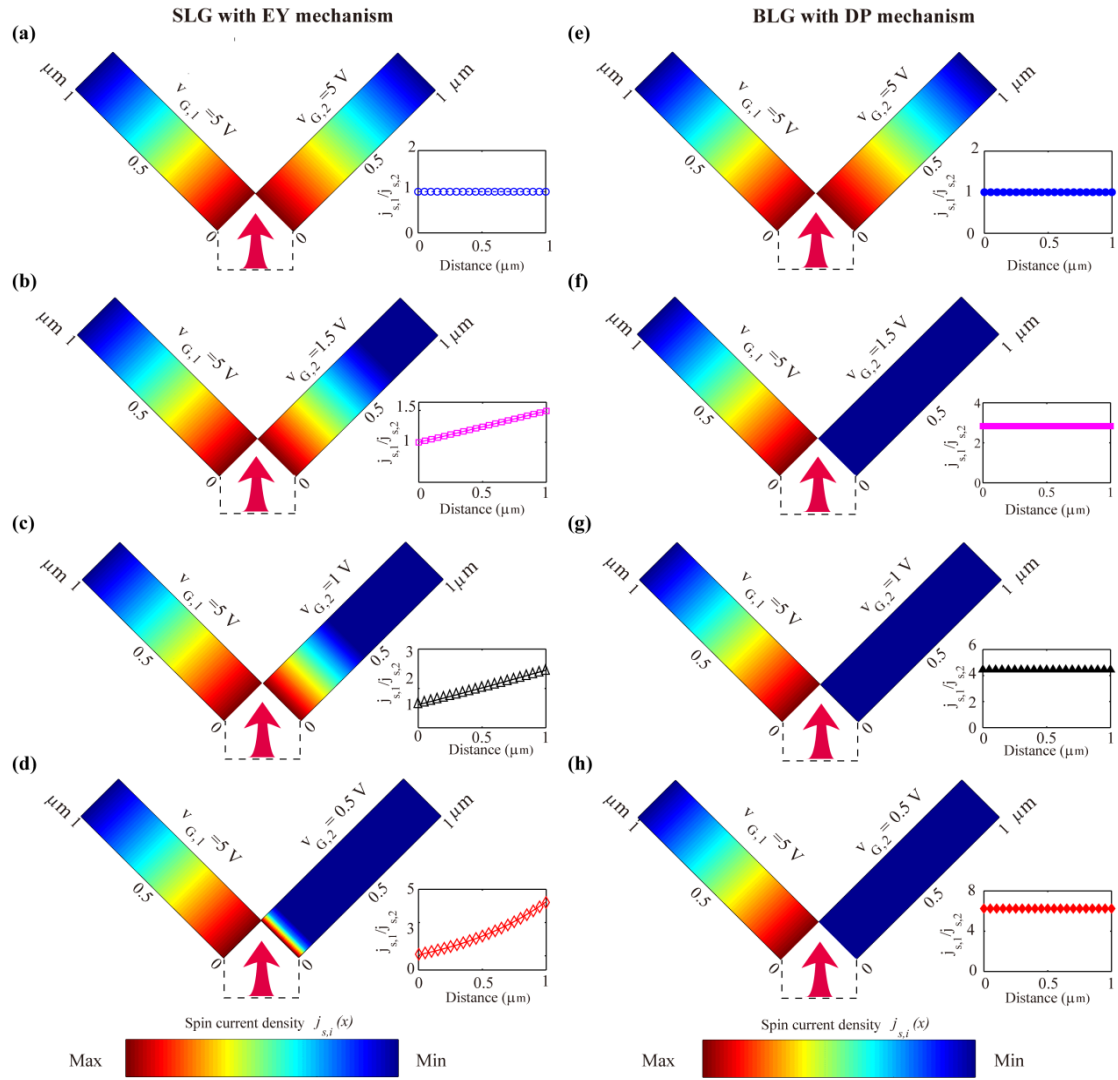

**Figure 3 | Device performance of Y-shaped GSDM.** Two typical examples of spin current manipulation by Y-shaped GSDM for SLG with EY relaxation mechanism and BLG with DP mechanism: two-dimensional mapping of the spin-current distribution at bifurcation ($x = 0$) and after propagation. The spin-current ratio $j_{s,1}(x)/j_{s,2}(x)$ as a function of distance $x$ is displayed in every insert figure. **(a)** and **(e)** The same voltage $v_{G,1} = v_{G,2} = 5$ V is applied for SLG with EY and BLG with DP, respectively. **(b)–(d)** Spin transport properties of SLG with EY mechanism with different gate voltages $v_{G,i}$. **(f)–(h)** Spin transport properties of BLG with DP mechanism using

different gate voltages $v_{G,i}$. Note that, the symmetric device geometry is considered, that is, $W_1 = W_2$; therefore, the spin current is represented directly by the spin current density as Eq. (1) in the earlier calculation. For the dielectric gate, $\varepsilon_G \approx 3.9$ and $t_G = 50$ nm.

**Table 1.** | **Expressions of the Fermi energy $E_F$, conductivity $\sigma$, momentum relaxation time $\tau_p$, spin relaxation time $\tau_s$, and spin diffusion length $\lambda_s$ as a function of carrier density n for SLG (left) or BLG (right) and for EY or DP spin relaxation mechanism.** The calculation in the text have been performed with $g_v = 2$ and $g_s = 2$ for respectively the valley and spin degeneracies, $v_F \approx 10^6$ ms$^{-1}$ for the Fermi velocity of SLG and $m^* \approx 0.033 m_e$ for effective mass of carriers in BLG.

|  | Single layer graphene (SLG) | | Bilayer graphene (BLG) | |
| --- | --- | --- | --- | --- |
| $E_F$ | $E_F = \hbar v_F \sqrt{\dfrac{4\pi n}{g_s g_v}}$ | | $E_F = \dfrac{2\pi \hbar^2 n}{m^* g_s g_v}$ | |
| $\sigma$ | $\sigma = \sigma(n) + \sigma_{min} = ne\mu + \sigma_{min}$ | | | |
| $\tau_p$ | $\tau_p^{SLG}(n) = \dfrac{\sigma}{e^2} \dfrac{2\pi \hbar}{v_F \sqrt{g_v g_s \pi n}}$ | | $\tau_p^{BLG}(n) = \dfrac{m^* \sigma}{e^2 n}$ | |
| spin relaxation | $\tau_{s,EY} \approx \dfrac{E_F^2}{\Delta_{EY}^2} \tau_p$ | $\tau_{s,DP} \approx \dfrac{\hbar^2}{4\Delta_{DP}^2} \dfrac{1}{\tau_p}$ | $\tau_{s,EY} \approx \dfrac{E_F^2}{\Delta_{EY}^2} \tau_p$ | $\tau_{s,DP} \approx \dfrac{\hbar^2}{4\Delta_{DP}^2} \dfrac{1}{\tau_p}$ |
| $\tau_s$ | $\dfrac{8\pi^{3/2} \hbar^3 v_F \sqrt{n}\sigma}{\Delta_{EY}^2 (g_v g_s)^{3/2} e^2}$ | $\dfrac{e^2 \hbar v_F \sqrt{g_v g_s}}{8\sqrt{\pi} \Delta_{DP}^2} \dfrac{\sqrt{n}}{\sigma}$ | $\dfrac{4\pi \hbar^4 n \sigma}{m^* \Delta_{EY}^2 (g_v g_s)^2 e^2}$ | $\dfrac{\hbar^2 e^2}{4\Delta_{DP}^2 m^*} \dfrac{n}{\sigma}$ |
| $\lambda_s$ | $\dfrac{\pi v_F \hbar^2 \sqrt{8}}{\Delta_{EY} g_v g_s} \sigma$ | $\dfrac{\hbar v_F}{\sqrt{8} \Delta_{DP}}$ | $\dfrac{2\pi \hbar^2 v_F}{\Delta_{EY} g_v g_s e^2} \sigma$ | $\dfrac{\hbar v_F}{\sqrt{8} \Delta_{DP}}$ |